\documentclass[11pt,a4paper]{article}
\usepackage{cite}
\usepackage{amsmath, amsthm,mathtools,url ,amssymb,upgreek,slashed}
\parskip=\baselineskip

\begin{document}
\begin{titlepage}
\begin{flushright}

\end{flushright}

\vskip 1.5in
\begin{center}
{\bf\Large{Searching for a Black Hole}}\vskip.3cm {\bf{\Large{ in the Outer Solar System}}}
\vskip
0.5cm { Edward Witten} \vskip 0.05in {\small{ \textit{Institute for Advanced Study}\vskip -.4cm
{\textit{Einstein Drive, Princeton, NJ 08540 USA}}}
}
\end{center}
\vskip 0.5in
\baselineskip 16pt
\begin{abstract}  There are hints of a novel object (``Planet 9'') with a mass $5-10$ $M_\oplus$ in the outer Solar System, at a distance of order 500 AU.
If it is a relatively conventional planet, it can be found in telescopic searches.   Alternatively, it has been suggested that this body might be a primordial black
hole (PBH).   In that case, conventional searches will fail.  A
possible alternative is to probe the gravitational field of this object using small, laser-launched spacecraft, like the ones envisioned in
 the Breakthrough Starshot
project.  With a velocity of order $.001~c$, such spacecraft can reach Planet 9 roughly a decade after launch and can discover it if they can report
 timing measurements accurate to $10^{-5}$ seconds back to Earth.
  \end{abstract}
\date{May, 2020}
\end{titlepage}
\def\km{{\mathrm{km}}}
\def\sec{{\mathrm{sec}}}
\def\be{\begin{equation}}
\def\ee{\end{equation}}

\def\AU{{\mathrm{AU}}}

Clustering of orbits of Kuiper belt objects has given a possible hint of the existence in the outer Solar System of a new body, dubbed Planet 9, with a mass
of roughly $5-10 \,\,M_\oplus$ and a distance from the Sun of order 500 AU \cite{TS,BB,BABB}.  Searches for this object are in progress.   On the
other hand, it has been suggested \cite{SU} that Planet 9 might really be a primordial black hole (PBH) or other exotic compact object.   As noted in \cite{SU}, gravitational microlensing observations
have bounded the cosmic abundance of objects in this mass range, finding a few events which might be due to free-floating planets
or to PBH's  with masses of roughly a few $M_\oplus$  \cite{OGLE1,OGLE}.  

If Planet 9 is really a PBH or other exotic compact object, conventional searches will come up short.  Under some assumptions about the nature of cosmic dark matter, 
there would be a detectable signal from annihilation events in a dark matter halo surrounding a PBH or other primordial compact object
\cite{SU}.   However, this is not guaranteed.  For example, if dark matter 
consists of QCD axions or even lighter scalar particles, a detectable annihilation signal would not be expected.   

One may also search for gravitational effects of Planet 9.   Indeed, tracking of Saturn's orbit via the Cassini spacecraft
has already provided a constraint on the parameters of Planet 9, along with a possible weak hint of its existence \cite{Cassini,Cassini2}. 

Future planetary missions might improve this constraint, but a spacecraft that could directly probe the environs of Planet 9 could potentially do much better.
There are two obvious difficulties:  Planet 9, if it exists, is very far away, and we do not know where it is.   On the first point, note that Voyager 1 was been the fastest
object leaving  the inner Solar System to date.  Having exited the region of the known planets
 with  a velocity of about 17 $\km/\sec$ (aided by gravity assists from Jupiter and
Saturn), it will reach 500 AU in about
150 years.   On the second point, at best we have a rough knowledge of the orbit of Planet 9 in space, without specific knowledge of where it currently is in its
orbit.   Thus a significant portion of the sky must be explored.   

Ideally, to search for Planet 9, one would like spacecraft velocities of (at least)  hundreds of kilometers per second, so as to reach 500 AU in a time of order a decade.
And one would like to launch hundreds of spacecraft (at least)  in different directions so that some would come within dozens of AU of Planet 9, rather than hundreds of
AU.   These conditions are well out of reach for conventional space missions, but they might be achievable by something along the lines of Breakthrough Starshot
 \cite{Starshot}.
This project aims to use powerful lasers to accelerate miniature (gram scale) spacecraft to mildly relativistic velocities ($\sim .2 c$) so as to reach nearby stars
in a couple of decades.  A useful description of parameters for Breakthrough Starshot is \cite{Parkin}.   That paper also describes a precursor
mission to explore the outer Solar System with a velocity of $.01 ~c$.   As we will see, in searching for Planet 9, a scaled down version with $v\sim .001~c$ might be preferable.   Reducing the velocity by a factor of 10 means that the spacecraft mass can be scaled up by a factor of 100, while keeping fixed the kinetic energy
that the spacecraft must reach.\footnote{However, the $.01~c$ mission sketched in \cite{Parkin} has a spacecraft mass of only $6.6$ mg, divided between the
sail and the payload.   Multiplying this by 100 while keeping the sail mass fixed leaves a payload mass of barely .65 grams.  More realistically, one should
repeat the analysis in \cite{Parkin} with the target mass required for the mission and a target velocity of order $.001~c$.}   

In the search for Planet 9, a project along the lines of Breakthrough Starshot has two major advantages.   Large velocities may be attainable, and it is practical to launch a very large number of
spacecraft, possibly hundreds or more.   The reason for the last statement is that there is a very large cost in building the acceleration system, but once it is available
it can be used multiple times comparatively inexpensively.  For example, in \cite{Parkin}, a rough cost estimate is given of $\$ 517$ million for the launch system of
the $.01~c$ project,  but the energy cost per launch is estimated at $\$8,000$.   Of course, a launch has other costs, including the cost of the spacecraft themselves.
Still it might be possible to launch hundreds or thousands of miniature spacecraft searching for Planet 9.  

To estimate the sensitivity that could be reached, consider
 a spacecraft incident on Planet 9 with velocity $v_0$ and impact parameter $\rho$.  If we ignore the gravity of Planet 9 as well as other perturbations,
the spacecraft travels on a straight line.    In a suitable coordinate system centered on Planet 9, with the
time coordinate $t$ defined so that $t=0$ at the moment of closest approach, the spacecraft
trajectory takes the form
\be\label{once}(x_0(t),y_0(t),z_0(t))=(v_0t,\rho,0). \ee
Taking into account the gravitational field of Planet 9, the $x$-component of the velocity of the probe will be $v_x(t)=v_0+v_1(t)$ where $v_1(t)$ is a small perturbation that
satisfies 
\be\label{ponce}\frac{d v_1(t)}{d t}=-\frac{GM x_0(t)}{(x_0(t)^2+\rho^2)^{3/2}}.\ee
Here $M$ is the mass of Planet 9.    Eqn. (\ref{once}) leads to $v_1(t)= GM/v_0(v_0^2t^2+\rho^2)^{1/2}$.    Integrating again, we find that the $x$-position
of the spacecraft can be approximated as $x(t)=v_0t+x_1(t)$ with
\be\label{xone} x_1(t)=\frac{GM}{v_0^2}\sinh^{-1}(v_0 t/\rho). \ee
This shift in position of the spacecraft will shift the arrival time on Earth of a signal from the spacecraft by\footnote{We assume that the spacecraft is moving
directly away from the Earth, which is a good approximation for a spacecraft launched from Earth with a velocity of hundreds of kilometers per second.}
\be\label{zone}\Delta\tau(t)=\frac{x_1(t)}{c}=\frac{GM}{v_0^2c} \sinh^{-1}(v_0t/\rho)\ee 
or
\be\label{plone}\Delta\tau(t)\cong 7\times 10^{-5}\,\,\mathrm{seconds}\cdot  \left(\frac{M}{5M_\oplus}\right)\left(\frac{10^{-3}\,\,c}{v_0}\right)^2 \sinh^{-1}(v_0t/\rho). \ee

This function increases for smaller $v_0$, but one
 does not want $v_0$ significantly less than $10^{-3}\,c$ because the mission will take too long.  (At $.001~c$, one reaches 500 AU in about eight years.)   
For reasonable values of $t$, the function $\sinh^{-1}(v_0t/\rho)$ will be of order 1, as we discuss momentarily.   So we see the basic parameters
required to search for Planet 9 via perturbations of the orbits of miniature spacecraft.   The spacecraft must be able to return time signals that
are accurate at least to $10^{-5}$ seconds, keeping under control at that level
all other perturbations (such as those due to the operation of the spacecraft or their interaction with
the dilute environment of the outer Solar System).   Sufficiently accurate timekeeping in a miniature spacecraft may be the biggest obstacle to this project, though there are numerous
other challenges.

What is an appropriate value of $\rho$?    Suppose that by studies of the Kuiper belt, Planet 9 is known to be located in a fraction $\varepsilon$ of the sky,
so that a solid angle $4\pi \varepsilon$ must be searched.   Suppose as well that it is practical to launch $N$ miniature spacecraft in directions uniformly
chosen in the  search region.   Also let $R$ be the distance to Planet 9.   The smallest impact parameters of all those spacecraft when they reach
Planet 9 will then be roughly $(4\pi\varepsilon / N)^{1/2}R$.   For example, if $\varepsilon=1/2$, $R=500~\AU$, and $N=1000$, this is
$\rho\cong 40$ AU.  In that case, if $v_0=.001~c$, then one year before (after) closest approach of the probe to Planet 9, $\sinh^{-1}(v_0t/\rho)$ is approximately $-1.2$
($+1.2$), so during that two year period $\Delta\tau$ changes by about $1.7\times 10^{-4}$ seconds.   Using a smaller value of $\varepsilon$ or $N$ would not change
this too much.  

It is interesting to compare these estimates with the Pioneer anomaly \cite{Pioneer}.
  That  anomaly was a discrepancy between the predicted and observed distances to the Pioneer spacecraft (which were launched in 1972 and 1973). It was
ultimately found to have a conventional explanation.   The effect considered here is several orders of magnitude smaller than the Pioneer anomaly,
and it has to be observed at a distance of order 500 AU, while the Pioneer anomaly showed up at a distance of $10-20$ AU.  Another point of comparison
is a proposal to test the gravitational force at a distance of order 100 AU using distance measurements to a conventional spacecraft \cite{G}.   After the initial
submission of this article to the arXiv, it was pointed out that timing of millisecond pulsars might possibly be sensitive enough to detect an object such as Planet 9
\cite{GLC}, and that the interstellar version of Planet 9 might be sensitive to the gravity of exoplanets \cite{CL}.    A variant of the proposal in the present paper
involves the transverse acceleration of miniature spacecraft, which might be observable via long baseline interferometry \cite{LR}.

If further study of the Kuiper belt strengthens the case for existence of Planet 9, but discovery via telescopic searches or  a dark matter annihilation signal does not
follow, then a direct search by a fleet of miniature spacecraft may become compelling.   Once Planet 9 is found by this
method, subsequent searches by the same method could pin down its location far more precisely and perhaps eventually make possible a close-up study of this
object.

\noindent{\it Acknowledgment}  I thank N. Arkani-Hamed  for discussions, and J. Scholtz and J. Unwin for
pointing out ref. \cite{G}.   Research  supported in part by  NSF Grant PHY-1911298.

\bibliographystyle{unsrt}

\end{document}